\documentclass[conference,a4paper]{IEEEtran}
\IEEEoverridecommandlockouts

\usepackage{array}
\usepackage{graphicx, float}
\usepackage{booktabs}
\usepackage{multirow}
\usepackage{adjustbox}
\usepackage{cite}
\usepackage{titlesec}
\usepackage{indentfirst}
\usepackage{amsmath,amssymb,amsfonts}
\usepackage{algorithmic}
\usepackage{algorithm}
\usepackage{comment}
\usepackage{graphicx}
\usepackage{textcomp}
\usepackage{xcolor}
\usepackage{array}    
\usepackage{longtable} 
\usepackage{xcolor} 
\usepackage{multirow} 
\def\BibTeX{{\rm B\kern-.05em{\sc i\kern-.025em b}\kern-.08em
    T\kern-.1667em\lower.7ex\hbox{E}\kern-.125emX}}
\usepackage{enumitem}
\setlist{nolistsep}
\usepackage{etoolbox}
\AfterEndEnvironment{enumerate}{\vskip-\lastskip}
\begin{document}


\title {Securing Critical Infrastructure in the AI Era: An Automated AI-Based Security Framework}

\title {Autonomous AI-Based Cybersecurity Framework for Critical Infrastructure: A Real-Time Threat Mitigation Approach}

\title{Autonomous AI-based Cybersecurity Framework for Critical Infrastructure: Real-Time Threat Mitigation}


\author{\IEEEauthorblockN{Jenifer Paulraj, Brindha Raghuraman, Nagarani Gopalakrishnan, Yazan Otoum}
\IEEEauthorblockA{School of Computer Science and Technology, Algoma University, Canada\\
Emails: \{jpaulraj, braghuraman, ngopalakrishnan, otoum\}@algomau.ca}
}

\maketitle

\IEEEpubidadjcol

\begin{abstract}

Critical infrastructure systems, including energy grids, healthcare facilities, transportation networks, and water distribution systems, are pivotal to societal stability and economic resilience. However, the increasing interconnectivity of these systems exposes them to various cyber threats, including ransomware, Denial-of-Service (DoS) attacks, and Advanced Persistent Threats (APTs). This paper examines cybersecurity vulnerabilities in critical infrastructure, highlighting the threat landscape, attack vectors, and the role of Artificial Intelligence (AI) in mitigating these risks. We propose a hybrid AI-driven cybersecurity framework to enhance real-time vulnerability detection, threat modelling, and automated remediation. This study also addresses the complexities of adversarial AI, regulatory compliance, and integration. Our findings provide actionable insights to strengthen the security and resilience of critical infrastructure systems against emerging cyber threats.


\end{abstract}


\section{Introduction}
\label{sec:introduction}

Critical Infrastructure (CI) systems, including energy grids, healthcare facilities \cite{otoum2024advancing}, water supplies, and transportation networks, form the foundation of modern society. With the growing integration of Industrial Control Systems (ICS), Internet of Things (IoT) devices, and cloud-based platforms, these systems face heightened exposure to sophisticated cyber threats. Notable incidents, such as the Colonial Pipeline ransomware attack and breaches into national power grids, underscore the severe consequences of cyberattacks on CI~\cite{maglaras2022cybersecurity, ojo2024aidriven, otoum2024enhancing}. The convergence of Operational Technology (OT) with Information Technology (IT) expands the attack surface, increasing susceptibility to zero-day vulnerabilities, ransomware, supply chain attacks, and Advanced Persistent Threats (APTs). Traditional IT-centric security mechanisms are no longer sufficient in these interconnected environments. Moreover, legacy systems in CI often lack modern security features, leaving critical assets vulnerable to attack. Recent advancements in Artificial Intelligence (AI) offer promising capabilities, such as real-time anomaly detection, predictive threat analytics, and automated incident response. However, existing commercial solutions, such as \textit{Darktrace} and \textit{CrowdStrike Falcon}, although AI-native, are often restricted to specific layers of defence and rely heavily on centralized management, making them difficult to scale or integrate across heterogeneous CI environments. Moreover, they do not fully automate the cybersecurity lifecycle nor utilize reinforcement learning for dynamic remediation planning. This study proposes an Autonomous AI-based Security Architecture (AISA), a novel, end-to-end framework that integrates AI-driven detection, prioritization, and reinforcement learning-powered remediation mapping, explicitly tailored for CI environments. Unlike traditional frameworks, AISA embeds automation throughout the entire cybersecurity lifecycle, from vulnerability scanning and threat prioritization to autonomous incident recovery. It further incorporates contextual scoring metrics such as CVSS Base Score, CVE Reference, and custom Impact Scores, with explicit design to compute, aggregate, and apply them in decision-making, unlike existing platforms that merely report them. AISA addresses key limitations identified in existing AI-based systems:
\begin{itemize}[leftmargin=1.5em]
    \item Most rely on predefined rule sets and human-in-the-loop remediation.
    \item Few utilize RL-based agents for generating adaptive mitigation strategies.
    \item No existing system offers full-cycle integration of real-time detection, prioritization, and autonomous recovery in CI-specific architectures.
\end{itemize}

To this end, AISA significantly reduces human dependency, accelerates response time, and supports compliance monitoring through automated regulatory alignment checks. Currently, tools like \textit{Splunk} and \textit{Nessus} assist in identifying vulnerabilities and recommending fixes, but they still require manual execution. In contrast, AISA closes this gap by implementing a fully autonomous response system that adapts to evolving threats while minimizing cost, complexity, and operational downtime. The main contributions of this paper are:
\begin{itemize}[leftmargin=1.5em]
    \item A critical analysis of cybersecurity risks across CI systems, spanning software, network, and hardware vulnerabilities.
    \item The design of a five-stage AI-enabled architecture that incorporates anomaly detection, threat scoring, reinforcement learning-based remediation mapping, and automated incident response.
    \item Evaluation of AISA’s impact in terms of improved threat response time, reduced system downtime, enhanced detection precision, and streamlined compliance reporting.
\end{itemize}

The remainder of this paper is organized as follows. Section II provides background on cybersecurity challenges in critical infrastructure systems. Section III reviews related work, outlining existing solutions and identifying key research gaps. In Section IV, we introduce the proposed AI-based security framework, AISA, detailing its architecture, operational workflow, and automation capabilities. Section V presents the evaluation results, demonstrating enhanced threat detection, accelerated response times, and reduced system downtime. Finally, Section VI concludes the paper with a summary of key findings and directions for future research.

\begin{figure*}[h]
    \centering
    \includegraphics[width=1\textwidth]{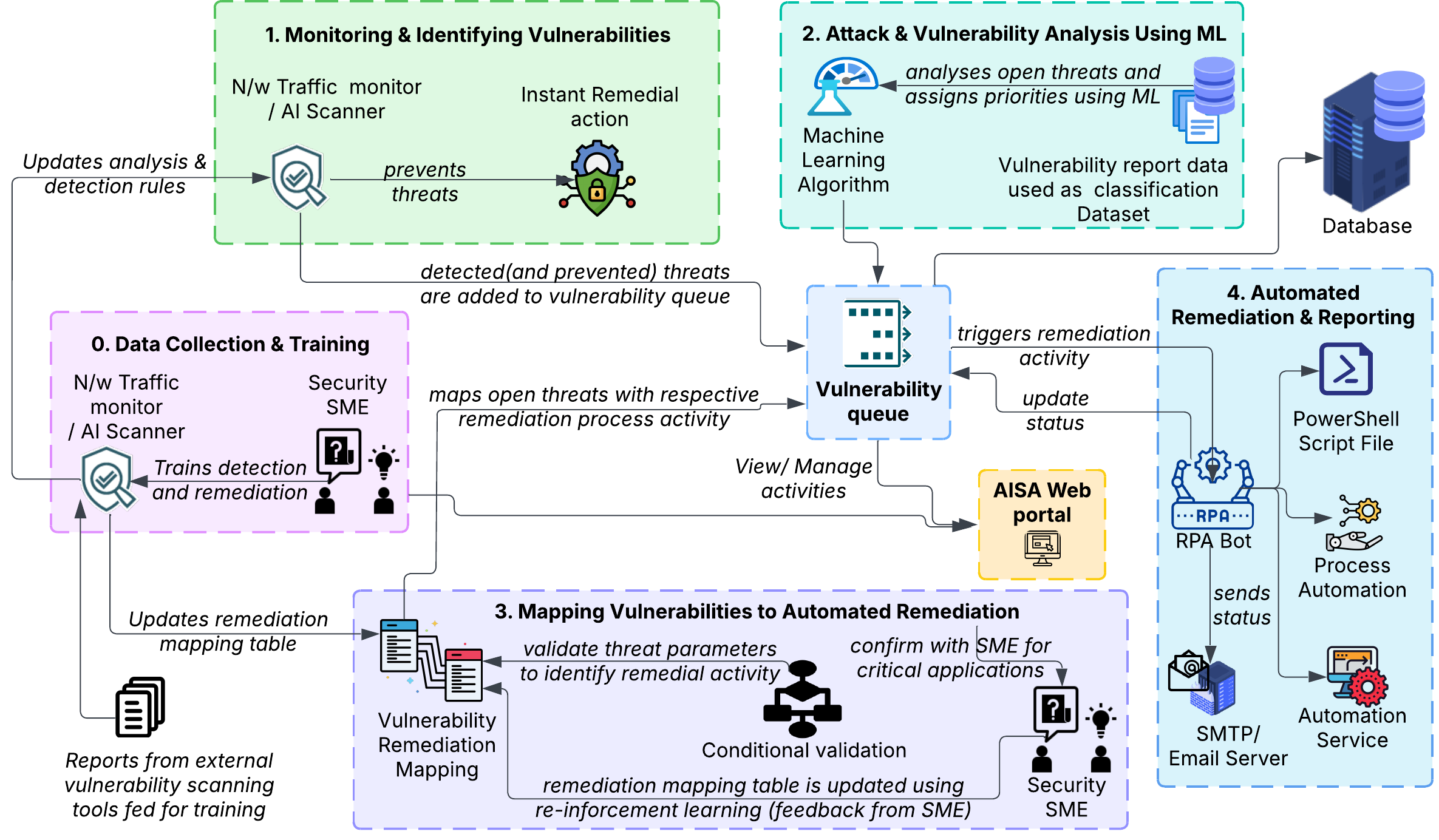} 
    \caption{AISA(AI Security Agent) Hyper Automation Solution - Design flow }
    \label{figuredesignflow}
\end{figure*}

\section{Background and Fundamental Concepts}\label{sec:background}

\begin{table*}[ht]
\centering
\renewcommand{\arraystretch}{1.2}
\scriptsize
\begin{tabular}{|p{0.07\textwidth}|p{0.09\textwidth}|p{0.04\textwidth}|p{0.08\textwidth}|p{0.05\textwidth}|p{0.15\textwidth}|p{0.15\textwidth}|p{0.15\textwidth}|} 
\hline
\textbf{Priority} & \textbf{Vulnerability} & \textbf{Impact Score} & \textbf{Key Attack Vector} & \textbf{CVE Reference} & \textbf{AI Analysis \& Detection} & \textbf{AI Automated Remediation} & \textbf{CVSS v3.1 Base Score} \\
\hline
\multirow{4}{*}{\textbf{High}} 
    & Unpatched Systems & 10 & Exploitable vulnerabilities (CVEs, zero-day attacks) & CVE-2024-21302 & AI scans infrastructure, cross-references CVE databases, and detects outdated software & AI auto-patches systems, applies virtual patching, and verifies fixes & Base Score: 10.0 (AV:N, AC:L, PR:N, UI:N, S:C, C:H, I:H, A:H) \\
    \cline{2-8}
    & Sophisticated Ransomware Attacks & 10 & Encrypts data, demands ransom & CVE-2024-6863 & AI detects file encryption anomalies, monitors process behaviors, and deploys honeypots & AI isolates infected systems, restores from backups, and blocks malicious traffic & Base Score: 10.0 (AV:N, AC:L, PR:N, UI:N, S:C, C:H, I:H, A:H) \\
    \cline{2-8}
    & Weak Passwords \& Authentication & 9.5 & Brute-force, credential stuffing, phishing & CVE-2024-3263 & AI scans stored passwords, analyzes user behavior, and monitors dark web credential leaks & AI enforces MFA, resets compromised passwords, and blocks unauthorized logins & Base Score: 9.5 (AV:N, AC:L, PR:L, UI:N, S:U, C:H, I:H, A:N) \\
    \cline{2-8}
    & DDoS Attacks & 9 & Overloads systems with malicious traffic & CVE-2024-39555 & AI monitors traffic patterns, detects botnet behavior, and differentiates attack traffic & AI rate-limits traffic, blocks malicious IPs, and scales resources dynamically & Base Score: 9.0 (AV:N, AC:L, PR:N, UI:N, S:U, C:N, I:N, A:H) \\
\hline
\multirow{4}{*}{\textbf{Medium-High}} 
    & Misconfigurations \& Default Settings & 8.5 & Default credentials, open ports & CVE-2024-36421 & AI scans network and cloud configurations, flags weak settings, and ensures compliance & AI auto-corrects misconfigurations, disables unused ports, and strengthens access controls & Base Score: 8.5 (AV:N, AC:L, PR:L, UI:N, S:U, C:H, I:H, A:L) \\
    \cline{2-8}
    & Advanced Persistent Threats (APTs) & 8.5 & Stealthy, long-term attacks & CVE-2024-12356 & AI detects lateral movement, privilege escalation, and spear-phishing attempts & AI isolates compromised accounts, enforces micro-segmentation, and blocks command-and-control (C2) traffic & Base Score: 8.5 (AV:N, AC:H, PR:L, UI:N, S:C, C:H, I:H, A:L) \\
    \cline{2-8}
    & Insider Threats & 8 & Privileged user misuse & CVE-2024-22169 & AI monitors access logs, flags unusual data transfers, and detects suspicious activities & AI dynamically adjusts privileges, restricts unauthorized actions, and flags anomalies in real-time & Base Score: 8.0 (AV:L, AC:L, PR:H, UI:N, S:C, C:H, I:H, A:L) \\
    \cline{2-8}
    & Improper Network Segmentation & 7.5 & Flat networks enable lateral movement & CVE-2024-43663 & AI analyzes traffic flow, detects excessive access permissions, and flags lateral movement & AI enforces micro-segmentation, restricting unauthorized access dynamically & Base Score: 7.5 (AV:A, AC:L, PR:L, UI:N, S:U, C:L, I:L, A:L) \\
\hline
\multirow{2}{*}{\textbf{Medium-Low}} 
    & Insecure Network Protocols & 7 & Weak legacy protocols (HTTP, FTP, Telnet) & CVE-2024-13872 & AI detects insecure protocols and outdated encryption standards & AI upgrades to secure protocols (TLS, SSH), blocks unsafe connections & Base Score: 7.0 (AV:N, AC:L, PR:N, UI:N, S:U, C:H, I:L, A:L) \\
    \cline{2-8}
    & Insufficient Logging \& Monitoring & 6.5 & Lack of detection capability & CVE-2024-54140 & AI ensures log collection, correlates security events, and detects anomalies & AI triggers real-time alerts, enforces automated incident response & Base Score: 6.5 (AV:L, AC:L, PR:L, UI:N, S:U, C:L, I:L, A:L) \\
\hline
\end{tabular}
\vspace{1em}
\caption{Scope of AI-Driven Cybersecurity Model – Vulnerabilities, Attack Vectors, CVEs, Detection, Remediation, and CVSS Base Scores}
\label{tab:cybersecurity_ai_analysis}
\end{table*}

Critical infrastructure security involves safeguarding essential services from cyber threats that could disrupt national security, public safety, and economic stability. CI sectors heavily depend on SCADA systems, ICS, and IoT-enabled components \cite{george2024cyber, sarker2024ai}. Modern CI systems operate in a highly connected digital ecosystem, relying on advanced communication protocols and distributed architectures. Integrating 5G networks, cloud computing, and blockchain technology has introduced new security paradigms while expanding the attack surface \cite{otoum2025blockchain}. These technologies enhance efficiency but also create vulnerabilities that adversaries can exploit through malware injections, insider threats, and coordinated cyber-physical attacks. A key challenge in securing CI is the convergence of IT and OT networks. While IT focuses on data security and network integrity, OT involves the real-time control of physical infrastructure, including power grids and water treatment facilities. This convergence exposes traditionally isolated OT systems to cyber risks, making them susceptible to remote exploitation, ransomware, and sabotage \cite{nist2018cybersecurity}. Security frameworks such as the NIST Cybersecurity Framework and ISO/IEC 27001 offer guidelines for risk management, while AI and ML-based solutions are emerging to enhance automation \cite{vigan2023cybersecurity}. Additionally, regulatory compliance frameworks, including the European Union’s Network and Information Security (NIS) Directive and the Cybersecurity Maturity Model Certification (CMMC), enforce security measures for CI operators to prevent cyber incidents \cite{bennouk2024comprehensive}. Traditional security tools, such as firewalls and Intrusion Detection Systems (IDS), are now complemented by AI-powered Security Orchestration, Automated Response (SOAR) platforms and anomaly detection models \cite{otoum2025llm}. AI-driven solutions provide enhanced threat detection, automated remediation, and predictive analytics to mitigate emerging risks. A hybrid model integrating AI, human oversight, and adaptive cybersecurity frameworks is necessary to strengthen CI resilience, ensuring a proactive defence mechanism that evolves with the threat landscape while maintaining regulatory compliance \cite{jimmy2021emerging}.

\section{State-of-the-Art Literature Review}\label{sec:literature}

Securing CI against evolving cyber threats has become increasingly complex as adversaries adopt sophisticated tactics. Recent studies highlight the increasing reliance on AI to enhance CI cybersecurity. AI-enhanced IDS improve threat visibility and response efficiency \cite{maglaras2022cybersecurity}. At the same time, Zero-Trust Architecture (ZTA) and AI-powered Security Information and Event Management (SIEM) platforms enhance real-time monitoring and policy enforcement in distributed environments \cite{salem2024advancing}. Advanced AI-driven techniques for vulnerability management and automated threat mitigation, especially in dynamic settings, are explored in \cite{tanikonda2022advanced}. While \cite{george2024cyber} emphasizes the urgency of proactive, cross-sector threat identification. The NIST Cybersecurity Framework supports the integration of AI into traditional risk management models \cite{nist2018cybersecurity}. Evolving adversarial AI threats are also discussed in \cite{vigan2023cybersecurity}. A comprehensive review of AI-enabled vulnerability detection proposes combining AI with blockchain for decentralized authentication \cite{bennouk2024comprehensive}, a growingly relevant approach as CI systems become increasingly interconnected. Finally, the synergy between AI and Robotic Process Automation (RPA) for scalable cybersecurity is discussed in \cite{kansal2024automation}, though challenges in securing dynamic, mission-critical CI environments remain underexplored. 
Key research gaps include:
\begin{itemize}
    \item Limited empirical evidence on AI’s real-time performance in detecting zero-day vulnerabilities.
    \item The need for standardized AI frameworks to ensure interoperability across different CI sectors.
    \item Lacks hybrid AI-human collaborative security approaches for enhanced situational awareness.
    \item Challenges applying blockchain for real-time threat intelligence while ensuring computational efficiency.
    \item Scalability concerns in integrating AI-powered cybersecurity measures within legacy CI infrastructures.
\end{itemize}

\section{Proposed Framework}\label{sec:proposedframework}

\subsection{Scope of the Solution}

Cybersecurity threats are evolving rapidly, with attackers using advanced techniques to exploit vulnerabilities in systems, networks, and authentication protocols. Traditional security measures often fall short in real-time threat detection and mitigation. AI significantly enhances cybersecurity by identifying vulnerabilities, detecting threats, and enabling automated remediation to reduce impact. This section classifies vulnerabilities by severity and impact score, highlighting how AI-based analysis and automation strengthen resilience. Table~\ref{tab:cybersecurity_ai_analysis} prioritizes network infrastructure threats, outlining impact scores and corresponding remediation steps.

\subsection{Proposed Model Workflow}

This study proposes an AI-driven cybersecurity model that integrates real-time anomaly detection, automated response, and regulatory compliance to enhance resilience against evolving threats. Tools like Splunk and Nessus help identify vulnerabilities and suggest remediation; however, current processes are resource-intensive, prone to errors, and expensive. Our approach aims to automate real-time prevention, detection, recovery, and remediation, minimizing impact while delivering financial and risk management benefits.

\begin{algorithm}
\caption{AI-Driven Remediation Script Execution}
\begin{algorithmic}[1]
\REQUIRE Identified vulnerability $V$
\IF{pre-existing remediation script exists for $V$}
    \STATE Execute the existing script
\ELSE
    \STATE AI analyzes vulnerability $V$, past remediations, and system configs
    \STATE AI generates a remediation script (PowerShell, Python, or RPA)
    \STATE Validate script using security policies
    \IF{$V$ affects a critical application}
        \STATE Send script to SME for approval; wait for approval
    \ENDIF
    \STATE Execute the validated (and if needed, approved) script
\ENDIF
\end{algorithmic}
\label{alg:RemediationSteps}
\end{algorithm}

\begin{figure*}
    \centering
      \includegraphics[width=0.3\linewidth]{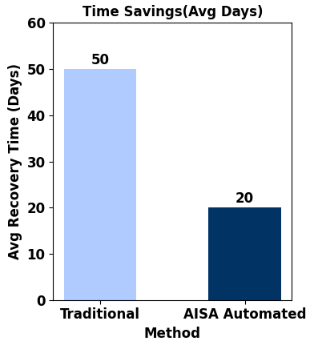}
     \includegraphics[width=0.3\linewidth]{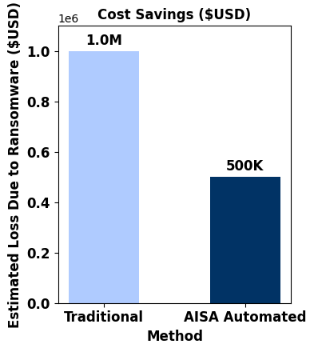}
     \includegraphics[width=0.3\linewidth]{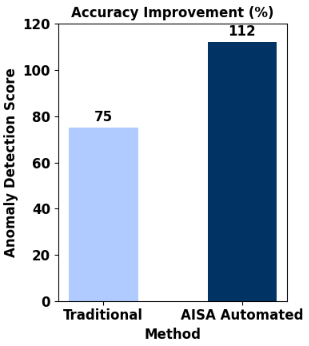}
    \caption{AISA vs Traditional - Savings and Accuracy}
    \label{figureSavings}
\end{figure*}

\begin{figure}
    \centering
    \includegraphics[width=0.5\textwidth]{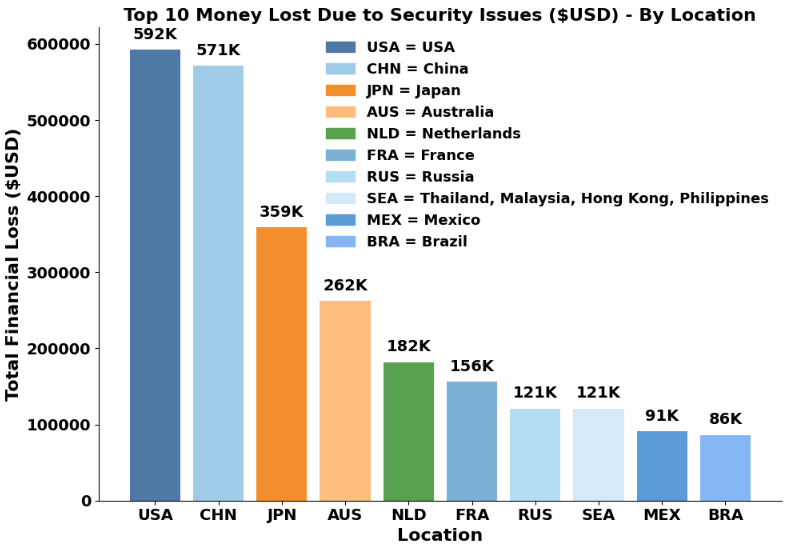}
    \caption{Financial Loss by Location - Ransomware issues}
    \label{figureLossByLocation}
\end{figure}
\begin{figure}
    \centering
     \includegraphics[width=0.5\textwidth]{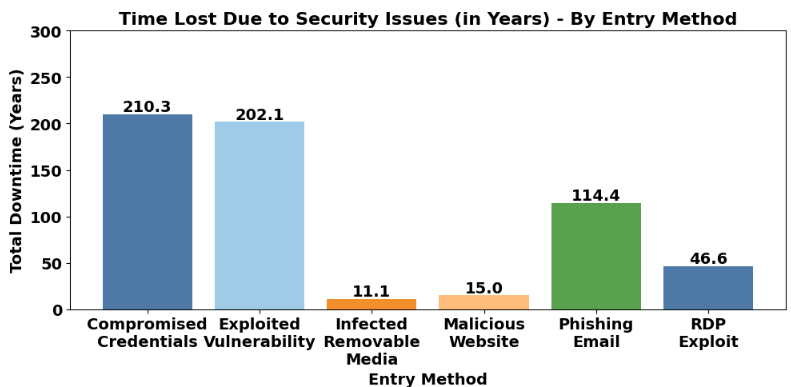} 
    \caption{Time Loss by Entry method - Ransomware issues}
    \label{figureLossByEntryMethod}
\end{figure}

\begin{figure}
        \centering
        \includegraphics[width=1.0\linewidth]{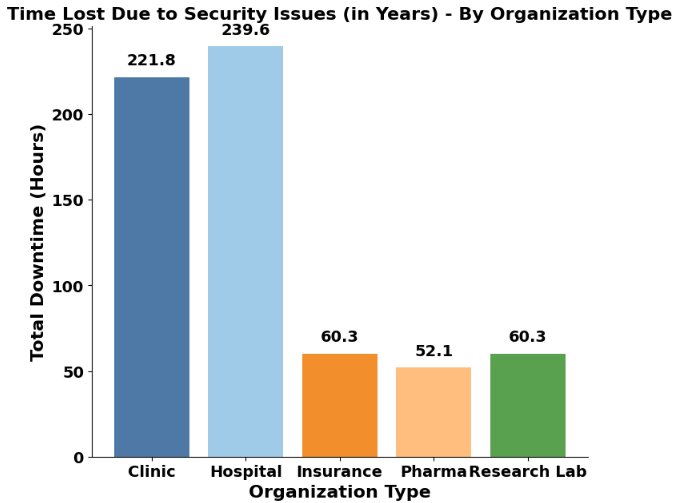}
        \caption{Ransomware Incidents - Time Lost By Org Type}
        \label{figureRansomwareIncidentsByOrgTypeandSize}
\end{figure}

\subsection{AI-Driven Threat Detection and Automated Remediation Framework}

The AISA automation solution follows a five-stage process to enhance cybersecurity through AI-driven threat detection, machine learning-based prioritization, and automated remediation. Fig.~\ref{figuredesignflow} represents the design flow of the end-to-end automation solution. It involves one setup stage and four execution stages. This automation solution consists of five key stages as follows:

\textbf{Stage 0: Data Collection \& Training (Setup Stage)} 

The foundation of the AISA automation system is established during Stage 0, a setup phase that involves training AI and machine learning models using historical cybersecurity data. This includes compiling attack records from SIEM systems, endpoint telemetry, threat intelligence feeds, and remediation outcomes from past incidents. AISA also integrates output from external vulnerability scanners, such as Nessus and Qualys, into its training data. Subject Matter Experts (SMEs) contribute domain knowledge to define remediation logic, policy constraints, and validation rules that guide the training process. A critical component of the training phase is the use of reinforcement learning (RL) to construct and refine AISA’s remediation mapping table. This structured knowledge base links specific vulnerabilities to optimal mitigation strategies. In this setting, AISA operates as an RL agent that learns through interactions with historical data and simulated remediation environments. Each remediation action taken in response to a given vulnerability scenario is evaluated through a reward signal, which reflects the effectiveness of the action in resolving the threat without introducing service disruption, compliance violations, or unintended side effects. Additionally, the reinforcement learning process is also augmented through SME involvement, whereby experts review and annotate the outcomes of specific remediation actions. These human-in-the-loop inputs serve as guided feedback that can override or reinforce machine-derived reward signals, ensuring that domain-specific nuances, regulatory constraints, and operational dependencies are encoded adequately into the learning loop. This hybrid feedback mechanism enhances both the accuracy and generalizability of AISA’s decision-making policy, particularly in complex or high-stakes environments where automated inference alone may be insufficient. For example, during implementation in a critical infrastructure environment, AISA was trained using five years of attack history involving operational technology assets such as PLCs and SCADA controllers. The training data included documented incidents of ransomware targeting ICS networks and zero-day exploits in outdated firmware. Vulnerability scanner outputs from Nessus were merged with SME-reviewed remediation actions to teach AISA how to correlate specific CVEs with effective mitigations. Using reinforcement learning, the system established mappings for over 3,000 CVEs. One of these, CVE-2024-21302, was associated with successful past mitigations that included firmware upgrades and network isolation. These insights formed the knowledge base used during live system execution.

\textbf{Stage 1: Monitoring \& Identifying Vulnerabilities}

In this stage, AISA continuously monitors network traffic, system logs, and endpoint behaviour using its AI Scanner. The scanner analyzes telemetry to detect suspicious behaviour, unauthorized access attempts, vulnerability exposures, and other indicators of compromise. It applies machine learning models to classify threats into low, medium, or high risk categories based on asset sensitivity, exposure levels, and anomaly severity. For high-risk threats, AISA may take immediate preventive actions such as isolating the device, restricting access, or generating alerts. Regardless of severity, all detected vulnerabilities are logged in a centralized queue for further analysis in the next stage. For example, during real-time network monitoring, AISA’s AI Scanner identifies abnormal outbound communication from a SCADA controller within a power grid substation. The scanner flags this as potentially malicious and inspects the firmware version, discovering it is running version X.0.2, which matches the known remote code execution vulnerability CVE-2024-21302. Recognizing the CVE as high-risk, AISA applies an instant network segmentation policy via a virtual firewall rule, isolating the SCADA controller from internet-facing interfaces. The system then logs the threat, including the associated CVE, the affected asset, the timestamp, and the initial mitigation action. This entry is added to the vulnerability queue, enabling the system to proceed with deeper threat analysis and prioritization.

\textbf{Stage 2: Attack \& Vulnerability Analysis Using ML} 

Once vulnerabilities are added to the queue, AISA performs deeper analysis using its AI Analyzer, which applies machine learning models trained on historical attack patterns and contextual risk factors. The Analyzer calculates a dynamic impact score for each vulnerability using factors such as CVSS severity, asset criticality, known exploit activity, dependency graphs, and environmental exposure. This score is more reflective of real-world business risk than static scoring systems. Based on the results, AISA generates a vulnerability report dataset and ranks threats, prioritizing high-impact issues for remediation first. For example, CVE-2024-21302, which was previously logged, is identified by AISA as an affected SCADA controller that manages multiple substations and serves as a critical node in the energy grid. Through dependency analysis, the system recognizes that several downstream systems rely on the controller’s data for operational continuity. Threat intelligence feeds show that the vulnerability is being actively exploited in targeted attacks. Taking all these variables into account, the AI Analyzer assigns the vulnerability a high dynamic impact score of 0.97. This elevated score ensures that the issue is prioritized at the top of the remediation queue for immediate attention.

\textbf{Stage 3: Mapping Vulnerabilities to Automated Remediation} 

In this stage, AISA maps each prioritized vulnerability to its appropriate remediation path using an AI-driven remediation mapper. This module references a continuously evolving remediation mapping table that captures historical remediation outcomes and reinforcement learning inputs. It determines whether full automation is possible or whether manual approval from a security expert is required. Conditional logic evaluates business criticality, system sensitivity, and policy thresholds to determine the appropriate response. If automation criteria are met, AISA prepares the remediation workflow for execution. If the system classifies the asset as business-critical, the remediation plan is submitted to a human SME through the AISA portal for review and approval. For example, for the high-priority vulnerability CVE-2024-21302, the AI Remediation Mapper reviews historical remediation data. It identifies a successful pattern that includes network isolation, firmware upgrade to version X.1.3, and a controller restart. Because the affected SCADA controller is classified as business-critical, AISA does not execute the plan immediately. Instead, it generates a detailed remediation workflow and routes it to an SME through the AISA portal. The SME validates the compatibility of the firmware version with the SCADA environment and confirms that the remediation process will not disrupt power delivery. Once approved, the workflow is scheduled for execution in the next stage.

\textbf{Stage 4: Automated Remediation \& Reporting} 

In the final execution stage, AISA carries out the approved remediation workflow using automation tools.  Algorithm~\ref{alg:RemediationSteps}  begins by checking whether a pre-existing remediation script is available for the identified vulnerability V. If such a script exists, it is executed immediately to provide a rapid and tested response. If no existing script is found, the AI analyzes the vulnerability by examining its nature, reviewing past remediation actions taken for similar issues, and considering the current system configurations. Based on this analysis, the AI generates a new remediation script tailored to the situation, selecting from formats such as PowerShell, Python, or RPA automation workflows depending on the environment and requirements. Before executing the script, it undergoes validation against predefined security policies to ensure compliance with organizational standards and minimize risks. If the vulnerability affects a critical application, the script is sent to a Subject Matter Expert (SME) for manual review and approval. The system then awaits this confirmation before proceeding. Once the script has been validated and, if necessary, approved by the SME, the algorithm executes the remediation to mitigate the vulnerability, thus completing the process. Depending on the asset configuration, the system executes predefined or AI-generated scripts through PowerShell and RPA bots to perform actions such as firmware patching, network reconfiguration, access control enforcement, and service restarts. The process includes integrity checks to ensure systems return to normal operation after remediation. The AISA web portal serves as a centralized management interface, enabling security teams and Subject Matter Experts to monitor remediation progress in real time, review workflow statuses, and intervene manually if necessary. This portal facilitates transparency and control over the automated remediation activities, ensuring critical decisions are visible and manageable by human operators. Throughout this stage, AISA sends real-time notifications to relevant stakeholders, including SOC leads and SMEs, to confirm the status of remediation tasks. All remediation actions, alerts, and outcomes are securely logged and compiled into compliance-ready reports aligned with standards such as ISO 27001, NIST CSF, or NERC CIP. These records close the security incident lifecycle, supporting auditability and continuous improvement. For example, in response to the approved remediation plan for CVE-2024-21302, AISA generates a custom PowerShell script that isolates the vulnerable SCADA controller, applies the firmware upgrade, and restarts necessary services. The web portal updates reflect these activities, allowing the SME to track progress and receive confirmation upon successful completion. An automated email alert is sent to the SOC lead and SME, and the vulnerability entry is marked resolved in the system queue, with a compliance report archived for future audits.

\section{Results and Analysis}\label{sec:results}

\begin{table*}[ht]
\centering
\renewcommand{\arraystretch}{1.3} 
\scriptsize 
\begin{tabular}{|p{0.24\textwidth}|c|c|c|c|}
\hline
\textbf{Metric} & \textbf{Traditional} & \textbf{AISA} & \textbf{Savings (\%)} & \textbf{Projected Savings} \\ \hline
\multicolumn{5}{|c|}{\textbf{Threat Response Time}} \\ \hline
Breach Containment Time (days) & 280 & 0.25 & 99.9 & 3 - 4M per breach \\ \hline
Patching Time (weeks) & 4 & 0.5 & 87.5 & - \\ \hline
DDoS Mitigation (Time) & - & - & - & - \\ \hline
Average Cost of a Data Breach (\$M) & 4.45 & - & - & 3 - 4M per breach \\ \hline
\multicolumn{5}{|c|}{\textbf{Detection \& Response Accuracy}} \\ \hline
Detection Accuracy for Critical Threats & 60 & 95 & 95\% Increase & 1 - 4M per year \\ \hline
False Positives & 30 & 2 & 98\% Reduction & - \\ \hline
Manual Intervention for Threat Response & 100 & 15 & 85\% Reduction & - \\ \hline
Potential Savings from Improved Accuracy & - & - & - & 1 - 4M per year \\ \hline
\multicolumn{5}{|c|}{\textbf{Business Continuity}} \\ \hline
Average Downtime per Cyberattack (days) & 21 & 0.5 & 97.6\% Reduction & 5 - 10M per year \\ \hline
Data Loss Reduction (\%) & 0 & 90 & 90\% & - \\ \hline
Uptime (\%) & 85 & 99.5 & 16.5\% Increase & - \\ \hline
\multicolumn{5}{|c|}{\textbf{Compliance \& Audit}} \\ \hline
Regulatory Risk Reduction (\%) & 0 & 85 & 85\% Reduction & 500K - 10M per breach \\ \hline
Compliance Standards (ISO, NIST, CIS) & Manual Checks & Automated & 100\% & - \\ \hline
Lower Insurance Premiums & - & Yes & - & - \\ \hline
\multicolumn{5}{|c|}{\textbf{Summary}} \\ \hline
Number of Breaches & 10 & 1 & 85\% Reduction & 10 - 20M per year \\ \hline
Incident Response Time (days) & 280 & 0.25 & 99.9\% & - \\ \hline
Human Intervention for Remediation (\%) & 100 & 25 & 75\% Reduction & - \\ \hline
\end{tabular}
\vspace{1em}
\caption{AISA vs. Traditional - Sample Results Projection}
\label{tab:aisa_vs_traditional}
\end{table*}

To analyze the benefits, we used ransomware datasets from Kaggle \cite{teamincribo_cybersecurity_2021,joebeachcapital_ransomware_2021,rivalytics_healthcare_ransomware_2021} to compare traditional approaches with AISA. Key metrics include time savings, cost savings, accuracy improvements, and financial losses from ransomware. The time savings in Fig.~\ref{figureSavings} are based on average recovery durations between manual and automated methods. Cost savings assume traditional recovery costs 10\% of revenue, while automation halves this. Accuracy, shown in Fig.~\ref{figureSavings}, reflects improvements in anomaly scores using automation. The Figures~\ref{figureLossByEntryMethod}, \ref{figureLossByLocation}, and \ref{figureRansomwareIncidentsByOrgTypeandSize} illustrate ransomware-related financial losses, grouped by entry method and region, highlighting the top 10 most affected countries. Assumptions follow industry benchmarks: a 60\% reduction in recovery time through automation, cost cuts from improved efficiency, and a 50\% boost in detection accuracy. Losses are further categorized by organization type, method of attack, and location to identify critical trends. Table~\ref{tab:aisa_vs_traditional} summarizes AISA’s benefits across four major dimensions. Reduction in Threat Response Time (Time-Based Impact) is enabled when AI reduces breach containment time by 99.9\%, from 280 days with traditional methods to under 15 minutes, resulting in \$3M–\$4M savings per breach \cite{IBM_2020_BreachReport}. Accuracy – Improving Detection and Response is also enhanced, with AI improving detection accuracy by 95\% and reducing false positives by 98\%, saving \$1M–\$4M annually \cite{McKinsey_2020_Cybersecurity, Gartner_2020_Security}. Impact on Business Continuity and Downtime Prevention is notable, as AI reduces average downtime from 21 days to just 0.5 days (97.6\% improvement), ensuring 99.5\% uptime and saving \$5M–\$10M annually \cite{Ponemon_Institute_2020_Downtime, IBM_2020_BreachReport}. Lastly, in the area of Compliance and Audit Efficiency – Reducing Regulatory Fines, AI improves compliance with standards like ISO 27001 and NIST, reducing regulatory risks by 85\% and mitigating potential fines of \$500K to \$10M per breach \cite{Deloitte_2020_Compliance}.


\section{CONCLUSION}
\label{sec:conclusion}

This paper addressed cybersecurity vulnerabilities in critical infrastructure, emphasizing the expanded attack surface resulting from the integration of IoT, cloud, and industrial control systems. AI-driven solutions are essential for enabling real-time monitoring, automated remediation, and proactive threat detection. While advancements in AI enhance security capabilities, challenges such as adversarial threats, IT/OT convergence, and regulatory fragmentation persist. Future efforts should prioritize the development of standardized, interoperable AI security frameworks and hybrid models that integrate intelligent automation with human oversight. Strengthening infrastructure resilience demands cross-sector collaboration to implement adaptive, self-healing, and potentially blockchain-enhanced security mechanisms.




\end{document}